# Quantum nature of a strongly-coupled single quantum dot-cavity system

K. Hennessy[1,2,*], A. Badolato[1*], M. Winger[1*], D. Gerace[1], M. Atatüre[1], S. Gulde[1], S. Fält[1], E. L. Hu[2], A. Imamoğlu[1]

[1]*Institute of Quantum Electronics, ETH Hönggerberg HPT G10, CH-8093 Zurich, Switzerland*

[2]*California NanoSystems Institute, University of California, Santa Barbara, California 93106*

[*]*These authors contributed equally to this work*

**Cavity quantum electrodynamics (QED) studies the interaction between a quantum emitter and a single radiation-field mode. When an atom is in strong coupling with a cavity mode[1,2], it is possible to realize key quantum information processing (QIP) tasks, such as controlled coherent coupling and entanglement of distinguishable quantum systems. Realizing these tasks in the solid state is clearly desirable, and coupling semiconductor self-assembled quantum dots (QDs) to monolithic optical cavities is a promising route to this end. However, validating the efficacy of QDs in QIP applications requires confirmation of the quantum nature of the QD-cavity system in the strong coupling regime. Here we find a confirmation by observing quantum correlations in photoluminescence (PL) from a photonic crystal (PC) nanocavity[3-5] interacting with one, and only one, QD located precisely at the cavity electric field maximum. When off-resonance, photon emission from the cavity mode and QD excitons is anti-correlated at the level of single quanta, proving that the mode is driven solely by the QD despite an energy mis-match between cavity and excitons. When tuned into resonance, the exciton and photon enter the strong-coupling regime of cavity-QED and the QD lifetime**

**reduces by a factor of 120. The photon stream from the cavity becomes anti-bunched, proving that the coupled exciton/photon system is in the quantum anharmonic regime. Our observations unequivocally show that QIP tasks requiring the quantum nonlinear regime are achievable in the solid state.**

Pursuit of solid-state cavity-QED is motivated by the possibility to fix the emitter location with respect to the mode electric field maximum and to enhance the emitter-cavity coupling by fabricating cavities with ultra-small volumes. Substantial progress has been made to this end, culminating in demonstration of strong coupling of a Cooper pair box to a superconducting transmission line microwave resonator[6] and of a QD to a nanoscale optical cavity mode[7-9]. However, further progress in QD cavity-QED has been partially hindered by the conventional practice of incorporating many QDs at random locations in the cavity, leading to indirect/off-resonant coupling to other QDs overlapping the mode. Recent experiments showing intense cavity emission, and even lasing, when the mode is non-resonant with the QDs[4,7-10] suggest that observation of quantum effects would require a nanocavity containing a single QD.

In ways similar to those used to position nanocavities to stacks of QDs[10], we were able to position with 30 nm accuracy a nanocavity to one, and only one, QD, aligning it to the electric-field maximum of the cavity mode as shown in Fig. 1a,b. This positioning technique offers the clear advantage of having only one QD interacting with the cavity, allowing us to study the coupled system with unprecedented clarity. Additionally, since the PC is not perturbed by other QDs, we achieve significantly higher $Q$s[11].

Our approach allows for pre-selection of QDs with desirable spectral properties. In the remainder of this report, we follow one such QD that exemplifies our single QD-cavity samples[12]. The measured spectrum (see Methods) of this QD (Fig 1c), consisted of a few narrow, isolated excitonic transitions that were promising for nanocavity coupling. We then fabricated[10,13] the PC with the lattice parameter *a* and hole radius *r/a* selected to lithographically tune the nanocavity mode to this precise spectral location. As shown in Fig. 1d, the cavity mode is spectrally positioned a few nanometers shorter in wavelength than the exciton transitions allowing us to tune the cavity by a thin-film condensation technique[14] and study the exciton-cavity interaction as a function of the detuning between mode and exciton $\Delta_\lambda = \lambda_x - \lambda_m$ ($\Delta_\omega = \omega_m - \omega_x$).

An important feature of the spectrum in Fig. 1d is that emission from the cavity mode is observed where there was clearly no QD emission prior to cavity fabrication. Cavity emission for arbitrary detunings challenges the notion that QDs may be fully described as artificial atoms with discrete energy levels. To explain this effect, recent reports speculate that mode emission is sustained by the coupling of a background continuum into the cavity, with the origin of the background attributed to the high-density of QDs in their samples[7] or to interaction with a continuum of higher order QD states[4]. In contrast to prior speculations, we find that even a single QD sustains efficient mode emission for all detunings observed in ~20 deterministically-coupled devices, from $-19 < \Delta_\lambda < 4.1$ nm. Since control cavities containing an InAs wetting layer but positively no QDs lack mode emission, we conclude that the coupling is mediated by the QD, as opposed to bulk or wetting layer states. We further investigate the nature of this coupling by measuring the quantum correlations between photons emitted from the exciton transitions and the spectrally detuned ($\Delta_\lambda = 4.1$ nm) cavity. This measurement corresponds to the second-order, normally-ordered, cross-correlation function $g^{(2)}_{x,m}(\tau) = \langle : I_x(t) I_m(t+\tau) : \rangle / \langle I_x(t) \rangle \langle I_m(t) \rangle$, where $I_{x(m)}$ refers to the intensity of the exciton (cavity-mode) photon stream. Arrival of a cavity (exciton) photon triggers a timer that stops upon detection of an exciton (cavity) photon for positive (negative) difference in arrival time (see Fig. 2a and Methods). We accumulate a histogram of the detection events (Fig. 2b). For zero time difference, we observe strong anti-bunching, a deviation from the uncorrelated case of $g^{(2)}(0)=1$ that directly proves the two emission events stem from the same single quantum emitter and are anti-correlated at the level of single quanta[15]. Remarkably, the time constant for negative time delay (8.5 ns) corresponds to the lifetime of the exciton $\tau_x = 7.6$ ns (Fig. 2c), while the time constant for positive time delay compares to the lifetime of the mode emission $\tau_m = 1.3$ ns (Fig. 2d). We rather would expect the lifetime of the states producing the stop pulse to dictate the respective time constant. Off-resonant cavity-exciton anti-correlation demonstrates the existence of a new, unidentified mechanism for channelling QD excitations into a non-resonant cavity mode, indicating a clear deviation from a simple artificial atom model of the QD.

Having studied the exciton-mode coupling for large detuning, we then tuned the mode into resonance with the exciton to study resonant interaction. Clear evidence of the strong coupling regime of cavity-QED is obtained from spectra recorded as a function of $\Delta_\lambda$. A subset of these spectra is plotted in Fig 3b, from which we note the following features: for $\Delta_\lambda > 0.05$ nm, the short-wavelength spectral feature remains cavity-like with a linewidth of $\Delta\lambda_- \approx \Delta\lambda_m = 0.071$ nm and the long-wavelength peak retains its excitonic nature with $\Delta\lambda_+ \approx \Delta\lambda_x = 0.025$ nm. For further decrease in $\Delta_\lambda$, the short-wavelength feature broadens and then separates into two distinct peaks. As tuning proceeds, the middle peak preserves exactly the wavelength and linewidth of the cavity mode[16] while the long- and short-wavelength peaks repel each other and assume equal linewidths of $\Delta\lambda_\pm \approx (\Delta\lambda_m + \Delta\lambda_x)/2 = 0.042$ nm. Our observation of a spectral triplet in the strong coupling regime is unique among recent reports of solid-state vacuum Rabi splitting[7-9]. The two outer peaks at $\Delta_\lambda = 0$ anti-cross as shown in Fig 3a and are identified as the polariton states of the strongly coupled exciton-photon system. The measured spectra excluding the center peak are compared to a calculated spectral function,

$$S(\omega) \propto \frac{A_+(\omega)}{(\omega-\Omega_+)^2 + \Gamma_+^2} + \frac{A_-(\omega)}{(\omega-\Omega_-)^2 + \Gamma_-^2} \qquad (1)$$

in which the amplitudes $A_\pm$ are slowly varying functions[17] of $\omega$ and the resonant frequencies $\Omega_\pm$ and linewidths $\Gamma_\pm$ are obtained from the predicted real and imaginary parts of the complex eigensolutions for the coupled system,

$$\Omega_\pm + i\Gamma_\pm = \frac{\omega_m + \omega_x}{2} - i\frac{\gamma_x + \gamma_m}{4} \pm \sqrt{g^2 + \frac{1}{4}\left(\Delta_\omega - i\frac{\gamma_x - \gamma_m}{2}\right)^2}. \qquad (2)$$

Here, $g$ is the exciton-mode coupling frequency estimated from the minimum observed polariton splitting $\Delta_\omega$ and the full-width at half-maximum (FWHM) of the mode $\gamma_m = 24.1\,\text{GHz} = 100\,\mu\text{eV}$ ($Q \approx 13,300$) and exciton $\gamma_x = 8.5\,\text{GHz} = 35\,\mu\text{eV}$. The calculated peak positions are plotted as the continuous lines in Fig 3a and show very good agreement with the measured ones. From this Rabi splitting trend, we $g = 18.4\,\text{GHz} = 76\,\mu\text{eV}$ and note that $g$ is reduced to ~70% of its maximum possible value due to the slight spatial mismatch between the QD and the electric field maximum of the cavity mode[18].

The predicted strong-coupling spectrum $S(\omega)$ accounts for two of the three peaks we observe experimentally—we attribute the additional peak to the pure photonic state of the cavity. A similar feature has been observed for Ba atoms strongly coupled to a high-finesse cavity[19] and is attributed to fluctuations in the number of atoms present in the cavity over time. Our data can be regarded as the solid-state analog of this scenario, in which fluctuations in emitter energy, rather than emitter number, occur over time. As the occupation of charging centers in the vicinity of the QD fluctuates, the exciton energy is renormalized via the Coulomb interaction and the detuning becomes large, $|\Delta_\omega| \gg g$. For such large detunings, the short-wavelength polariton reverts to the pure photonic state of the cavity. It is plausible that the QD exciton emission in this case leads to one of the longer-wavelength PL lines depicted in Fig.1d. In our experiments, we collect a time-average of the two regimes resulting in the observed spectral triplet[20].

The clear signatures of strong coupling in Fig. 3 are confirmed by measuring $\tau_x$ as the mode is tuned into resonance with the exciton. Modification of $\tau_x$ is predicted to be particularly strong in a PC environment due to the depleted density of optical states for energies in the photonic bandgap[21], and confirmed by our measurement at large $\Delta_\lambda = 4.1\,\text{nm}$ of long $\tau_x = 7.6\,\text{ns}$ in comparison to the lifetime in bulk material $\tau_0 \approx 1\,\text{ns}$. On the other hand, when the cavity mode is spectrally near the exciton, $\tau_x$ should be greatly reduced since the optical density of states at $\lambda_m$ increases over the bulk value by a factor proportional to $Q/V$[22,23]. As the cavity is tuned into resonance, $\tau_x$ decreases rapidly to $\tau_x = 1.6\,\text{ns}$ at $\Delta_\lambda = 1.26\,\text{nm}$ and further diminishes to a carrier-capture limited $\tau_x = 60$ ps at $\Delta_\lambda = 0$ ( Fig. 4a,b). The reduction in lifetime by a factor of 120 confirms that our QD is coupled to the cavity mode and spatially located near the cavity center. Such short lifetimes, approaching the cavity photon storage time, support the spectral evidence that the exciton-photon coupling $g$ is sufficiently large for the system to be in the strong coupling regime. In order to estimate $g$ independently from the Rabi splitting, we assume that $\tau_x$ is primarily determined by a Lorentzian dependence on the detuning[24] and fit the experimental data (see Methods) to obtain $g = 20.7\,\text{GHz} = 86$ µeV which agrees very well with the value obtained from the Rabi splitting.

The spectral anti-crossing we observe is unequivocally a result of strong coupling; however, we emphasize that the two-Lorentzian lineshape Eq. (1) also describes exactly two coupled classical harmonic oscillators[25]. Therefore, Rabi splitting alone is not sufficient to discriminate between a quantum nonlinear regime arising from single exciton-photon interaction and a classical regime describing the coupling behaviour of two classical oscillators. For cavity photon occupation numbers above one, the quantum nonlinear regime is manifested in the Jaynes-Cummings ladder spectrum[17], which is absent in the case of two strongly coupled classical oscillators. A direct experimental demonstration of the quantum nonlinear regime has so far only been reported for single atoms flying through microwave cavities[26]. Recently, the quantum nonlinear regime has been inferred by monitoring the statistics of the photon stream emitted by an optical cavity[27]. In QD cavity-QED, in which the artificial atom is comprised of many atoms embedded in a host matrix, it is crucial to confirm that the system manifests true quantum behaviour. To this end, and along the lines of experiments with atomic systems[27], we measure the second-order auto-correlation function $g_{m,m}^{(2)}(\tau)$ of the cavity photon stream under pulsed excitation. For $\Delta_\lambda = 4.1$ nm, the cavity displays nearly Poissonian photon statistics as shown in Fig. 4c, with $g_{m,m}^{(2)}(0) \approx 1$, revealing its harmonicity. That is, the cavity can accept multiple photons at the same time. As we tune the system into strong coupling, the statistics of the cavity photon stream dramatically change to sub-Poissonian (Fig. 4d), with a central peak area at $\tau = 0$ that is only 54% of the average peak area at other times. We note that the area of the peak at $\tau = 0$ is increased from the ideal value of zero by two mechanisms. First, the central peak in the observed spectral triplet, which is *not* significantly anti-bunched, contributes 45% of the total collected PL as determined by the three-Lorentzian fit shown in Fig. 3b. Second, in strong coupling, the carrier capture process proceeds on time scales longer than the polariton decay time (13.3 ps) so that it is possible for the system to undergo multiple capture/emission events per excitation pulse. With these factors in mind, the measured $g_{m,m}^{(2)}(\tau)$ proves clearly that the quantum anharmonic regime is achievable in the solid-state.

We note that the observed anti-bunching in $g_{m,m}^{(2)}(\tau)$ is the solid-state analog of the photon-blockade behavior reported in atomic cavity-QED[27]. The use of reflection/transmission measurements carried out when the QD and the cavity mode are on resonance could be used to demonstrate the photon blockade effect[28] in a more direct manner, reaching the ultimate limit of solid-state nonlinear optics at the single-photon level. The results of this work encourage the experimental pursuit of QIP tasks in the solid state, such as the deterministic coupling of two QDs mediated by a common cavity mode[29].

**METHODS**

**Semiconductor material.** The semiconductor heterostructure used for subsequent PC fabrication was grown on a (100) semi-insulating GaAs substrate by molecular beam epitaxy (MBE). The epitaxial structure consists of a 126 nm GaAs slab incorporating a layer of InAs quantum dots at the center with a density ranging from zero to $3/\mu m^2$ across the sample. By annealing the quantum dots in-situ when partially capped with GaAs, we tuned the quantum-dot ground state energies to ~1.3 eV. The PC slab is grown above a 1 µm $Al_{0.7}Ga_{0.3}As$ sacrificial layer that is later removed to provide vertical optical isolation.

**Optical chacterization.** For micro-PL experiments, samples are mounted in a liquid-helium flow cryostat operating at 4.2 K. Excitation is provided by an above band-gap ($\lambda$ = 780 nm) continuous wave diode laser focused to a 1µm spot using a 50× microscope objective (NA = 0.55). The PL signal is collected by the same microscope objective and sent to a 750 mm grating spectrograph. Dispersion on a holographic grating with 1500 grooves per millimeter yields a spectral resolution of 21 pm, and light is detected with a liquid-nitrogen cooled charge coupled device camera.

In time dependent photoluminescence for the determination of lifetimes, we excite the sample with a *Q*-switch pulsed diode laser at a repetition rate of either 40MHz or 80MHz. We spectrally filter the PL using the spectrometer and focus it on an avalanche photodiode (APD) single photon counting module. A time to amplitude converter (TAC) transforms time differences between photon detection events and synchronization pulses provided by the laser to electrical signals of corresponding amplitude. A multichannel analyzer accumulates a histogram of the measured intervals.

We can choose between two different types of APDs to obtain time resolution of either ~70 ps or ~400 ps, with the latter having significantly higher quantum efficiency.

In auto-correlation measurements, filtered light is sent to a Hanbury-Brown-Twiss interferometer, consisting of a 50:50 plate beam splitter with an APD at either output. Time-correlation of the signals is done in a manner similar to that in lifetime measurements. Cross-correlation measurements are done in the same way after the two wavelengths of interest are spatially separated by the spectrometer.

**Estimation of *g*.** We assume that $\tau_x$ is primarily determined by a Lorentzian dependence on the detuning[24] and fit the experimental data with the relation $\hbar/\tau_x = \gamma_{tot} = \gamma_b + \gamma_{SE}$, in which $\gamma_b$ is the background emission rate into all other modes and $\gamma_{SE} = \gamma_m g^2 /[\Delta_\omega^2 + (\gamma_m/2)^2]$ is the spontaneous emission rate into the cavity mode. The best fit is given by $\gamma_b = 15$ MHz $= 0.06$ μeV and $g = 20.7$ GHz $= 86$ μeV. The first value corresponds to a lifetime of 11 ns, strongly modified from the bulk lifetime by the photonic bandgap. The second value confirms that in our system the strong coupling condition $g^2 > (\gamma_x - \gamma_m)^2/16$ is satisfied and agrees very well with the value of *g* estimated from the vacuum-field Rabi splitting.

**Correspondence and requests for materials should be addressed to A.I. (imamoglu@phys.ethz.ch).**

**Figure Captions**

**Figure 1** Positioning a PC cavity mode to a single buried QD. **a**, AFM topography of a PC nanocavity with a single QD. **b**, Electric field intensity of the PC cavity mode showing that the location of the buried QD, indicated by the teal dot, overlaps the field maximum. **c**, PL spectrum before cavity fabrication of a single QD selected for cavity coupling based on clear emission from a few of discrete excitonic transitions. **d**, PL spectrum from the same QD after cavity fabrication showing emission from the cavity at 942.5 nm.

**Figure 2** Cross-correlation histogram and time-resolved PL from the QD-cavity system with $\Delta_\lambda = 4.1$ nm. **a**, Schematic of the optical system used in cross-correlation measurements. The mode and exciton wavelengths are separated by a grating monochromator and imaged on separate single-photon detectors. **b**, Cross-correlation histogram of the mode and exciton. The two emission events are anti-correlated at the single-photon level, giving rise to an anti-bunching in $g^{(2)}(\tau)$. **c**, The exciton lifetime $\tau_\mathrm{x} = 7.6$ ns compares to the time constant observed in the cross-correlation histogram for negative time differences. **d**, The lifetime of the mode emission $\tau_\mathrm{m} = 1.3$ ns compares to the time constant observed in the cross-correlation histogram for positive time differences.

**Figure 3** Characteristics of the strong coupling regime in the spectral domain. **a**, Wavelength of the polaritons for various detunings $\Delta_\lambda$. Calculated peak positions of the spectral function $S(\omega)$ describing the strongly coupled system are plotted in solid lines with measured peak positions extracted from PL plotted in red and blue dots. **b**, Spectra of the two anti-crossing polariton states near zero detuning. An additional peak is identified as the pure photonic state of the cavity.

**Figure 4** Characteristics of the strong coupling regime in the time domain. **a**, The exciton lifetime $\tau_x$ is reduced to 1.6 ns when the detuning is $\Delta_\lambda = 1.3$ nm. **b**, At exact spectral resonance, $\tau_x$ decreases to 60 ps, a reduction by a factor of 120 from the value at $\Delta_\lambda = 4.1$ nm. **c**, Autocorrelation function of the cavity mode off-resonance with the exciton, showing no significant quantum correlations. **d**, Autocorrelation of the strongly-coupled cavity-QD system demonstrating strong quantum correlations in the form of photon anti-bunching, where the central peak area is 54% of the area of peaks at other times.

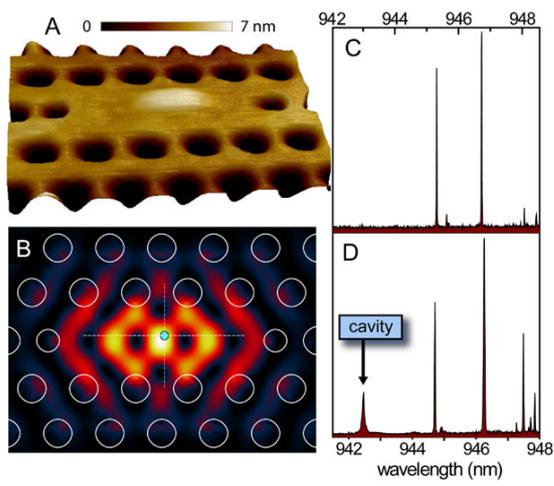

Figure 1

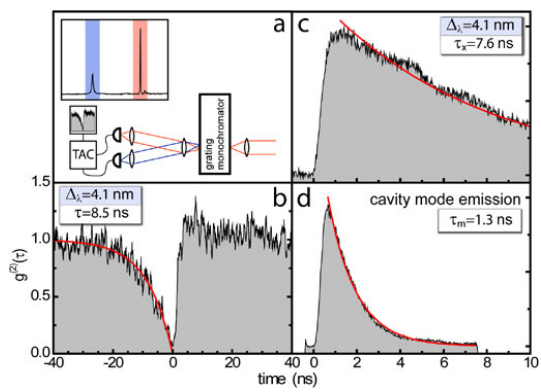

Figure 2

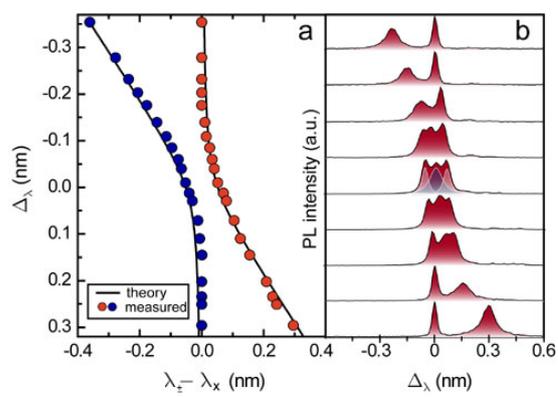

Figure 3

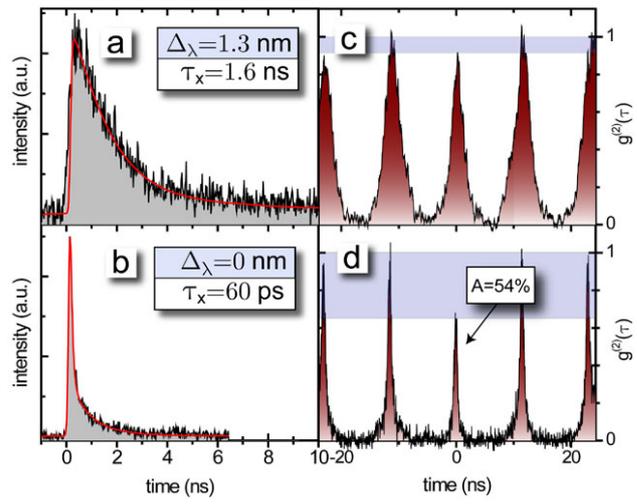

Figure 4